\documentclass[twocolumn,preprintnumbers,superscriptaddress,nofootinbib,aps,prd,floatfix]{revtex4}
\usepackage{amsmath,amssymb}
\usepackage{graphicx} %loads the graphicx.sty package 
\usepackage{epstopdf} %loads the epstopdf.sty package 
\usepackage{slashed}
\usepackage{subfigure}
\usepackage{color}
\usepackage{multirow,footmisc}
\usepackage{hyperref}

\usepackage{footmisc}

\usepackage{array}
\newcolumntype{L}[1]{>{\raggedright\let\newline\\\arraybackslash\hspace{0pt}}m{#1}}
\newcolumntype{C}[1]{>{\centering\let\newline\\\arraybackslash\hspace{0pt}}m{#1}}
\newcolumntype{R}[1]{>{\raggedleft\let\newline\\\arraybackslash\hspace{0pt}}m{#1}}

\newcommand{\be}{\begin{eqnarray*}}
\newcommand{\ee}{\end{eqnarray*}}

\newcommand{\bee}{\begin{eqnarray}}
\newcommand{\eee}{\end{eqnarray}}
\newcommand{\beeq}{\begin{equation}}
\newcommand{\eeeq}{\end{equation}}

\newcommand{\tomone}{$\times 10^{-1}$}
\newcommand{\tomtwo}{$\times 10^{-2}$}
\newcommand{\tomthree}{$\times 10^{-3}$}
\newcommand{\tomfour}{$\times 10^{-4}$}
\newcommand{\tomfive}{$\times 10^{-5}$}

\newcommand{\toone}{$\times 10^{1}$}
\newcommand{\totwo}{$\times 10^{2}$}

\begin{document}

%%%%%%%%%%%%%%%%%%%%%%%%%%%%%%%%%%%%%%%%%%%%%%%%%%%%%%%%%%
\title{Higgs Self-Coupling Measurements at a 100 TeV Hadron Collider}
%%%%%%%%%%%%%%%%%%%%%%%%%%%%%%%%%%%%%%%%%%%%%%%%%%%%%%%%%%
%
%

\begin{abstract}
  An important physics goal of a possible next-generation high-energy
  hadron collider will be precision characterisation of the Higgs
  sector and electroweak symmetry breaking. A crucial part of
  understanding the nature of electroweak symmetry breaking is
  measuring the Higgs self-interactions. We study dihiggs production
  in proton-proton collisions at 100~TeV centre of mass energy in
  order to estimate the sensitivity such a machine would have to
  variations in the trilinear Higgs coupling around the Standard Model
  expectation. We focus on the $b\bar b \gamma \gamma$ final state,
  including possible enhancements in sensitivity by exploiting dihiggs
  recoils against a hard jet. We find that it should be possible to
  measure the trilinear self-coupling with 40\% accuracy given 3/ab
  and 12\% with 30/ab of data.
\end{abstract}
%%%%%%%%%%%%%%%%%%%%%%%%%%%%%%%%%%%%%%%%%%%%%%%%%%%%%%%%%%

\author{Alan J. Barr} %\email{}
\affiliation{Denys Wilkinson Building, Department of Physics,\\ Oxford, OX1 3RH, UK\\[0.1cm]}
\author{Matthew J. Dolan} %\email{}
\affiliation{Theory Group, SLAC National Accelerator Laboratory,\\Menlo Park, CA 94025, USA\\[0.1cm]}
\author{Christoph Englert} %\email{christoph.englert@glasgow.ac.uk}
\affiliation{SUPA, School of Physics and Astronomy, University of
  Glasgow,\\Glasgow, G12 8QQ, UK\\[0.1cm]}
\author{Danilo Enoque Ferreira de Lima} %\email{}
\affiliation{SUPA, School of Physics and Astronomy, University of
  Glasgow,\\Glasgow, G12 8QQ, UK\\[0.1cm]}
\affiliation{Institute for Particle Physics Phenomenology, Department
  of Physics,\\Durham University, DH1 3LE, UK\\[0.1cm]} 
\author{Michael Spannowsky} %\email{michael.spannowsky@durham.ac.uk}
\affiliation{Institute for Particle Physics Phenomenology, Department
  of Physics,\\Durham University, DH1 3LE, UK\\[0.1cm]}
%

%\emailAdd{A.Barr1@physics.ox.ac.uk}
%\emailAdd{mdolan@slac.stanford.edu}
%\emailAdd{christoph.englert@glasgow.ac.uk}
%\emailAdd{danilo.ferreiradelima@glasgow.ac.uk}
%\emailAdd{michael.spannowsky@durham.ac.uk}

%\pacs{}
\preprint{IPPP/14/110, DCPT/14/220}

%\twocolumn[{
%\begin{@twocolumnfalse}
\maketitle
%\end{@twocolumnfalse}
%\flushbottom
%}]

%%%%%%%%%%%%%%%%%%%%%%%%%%%%%%%%%%%%%%%%%%%%%%%%%%%%%%%%%%
\section{Introduction}
\label{sec:intro}
%%%%%%%%%%%%%%%%%%%%%%%%%%%%%%%%%%%%%%%%5

The discovery of the Higgs boson~\cite{orig} at the Large Hadron
Collider (LHC)~\cite{discovery} has led to an extensive experimental
and theoretical effort to measure and constrain its properties in
order to understand in detail the mechanism of electroweak symmetry
breaking (EWSB)~\cite{Hcoup}. A crucial diagnostic in this process is
the measurement of the Higgs self-couplings, which directly probe the
higher order structure of the Higgs potential and BSM
effects~\cite{bsm,contino,Goertz:2014qta}.  While measurement of the
quartic Higgs coupling seems unlikely to be possible at any realistic
future hadron collider~\cite{rauch}, constraints can be set on the
Higgs trilinear coupling $\lambda$  by studying dihiggs
production~\cite{nigel}. In the Standard Model (SM) the coupling
$\lambda$ can be expressed in terms of the fundamental SM Lagrangian
parameters
\begin{multline}
  \label{eq:higgspot}
  V(H^\dagger H) =
  \mu^2 H^\dagger H + \eta (H^\dagger H)^2 \\
  \longrightarrow {1\over 2} m_h^2h^2 + \sqrt{ {\eta\over 2}}m_h h^3 +
  {\eta\over 4}h^4\,
\end{multline}
where we have expanded the potential around the Higgs vacuum
expectation value (vev), such that
$\lambda_{\text{SM}}=m_h\sqrt{\eta/2}$.

Research into dihiggs phenomenology has undergone a renaissance since
the Higgs discovery at the LHC. Well studied final states in the gluon
fusion production mode now include $b\bar b \tau\tau$~\cite{us,us2,baur68},
$b\bar b WW$~\cite{us,Papaefstathiou:2012qe} and $b\bar b b \bar
b$~\cite{us,deLima:2014dta,baur68}. There has also been significant work on
the vector boson fusion (VBF)~\cite{Dolan:2013rja} and $t\bar t h h
$~\cite{Englert:2014uqa, Liu:2014rva} production mechanisms. However,
due to it being the dominant production mechanism we focus exclusively
on gluon fusion in this article.

Early work on measuring Higgs trilinears at the LHC
includes~\cite{bauretal}, which suggested the $b\bar b \gamma\gamma$
final state as a promising possibility. While recent studies by
theoreticians generally agree with the results of that
article~\cite{Barger:2013jfa,Yao:2013ika}, evaluations from the
ATLAS~\cite{atlasecfa} and CMS collaborations~\cite{cmsecfa} find that
dihiggs production can be measured with considerably lower
significance than previously quoted (1.3 and $2~\sigma$ respectively
after $3000/\text{fb}$), corresponding in the ATLAS analysis to an
allowed range of $8.7 \geq \lambda /\lambda_{\text{SM}} \geq -1.3$ for
the Higgs trilinear coupling. This discrepancy between theorists and
experimentalists simulations is due to the treatment of backgrounds
which are due to fakes: either light jets faking photons or light jets
faking $b$-jets. A reliable estimate of the fake rate for various
reconstructed physics objects is thus a crucial component of any
analysis in this channel.

%%%%%%%%%%%%%%%%%%%%%%%%%%%%%%%%%%%
\begin{figure*}[!t]
  \centering
  \subfigure[]{\includegraphics[width=0.47\textwidth]{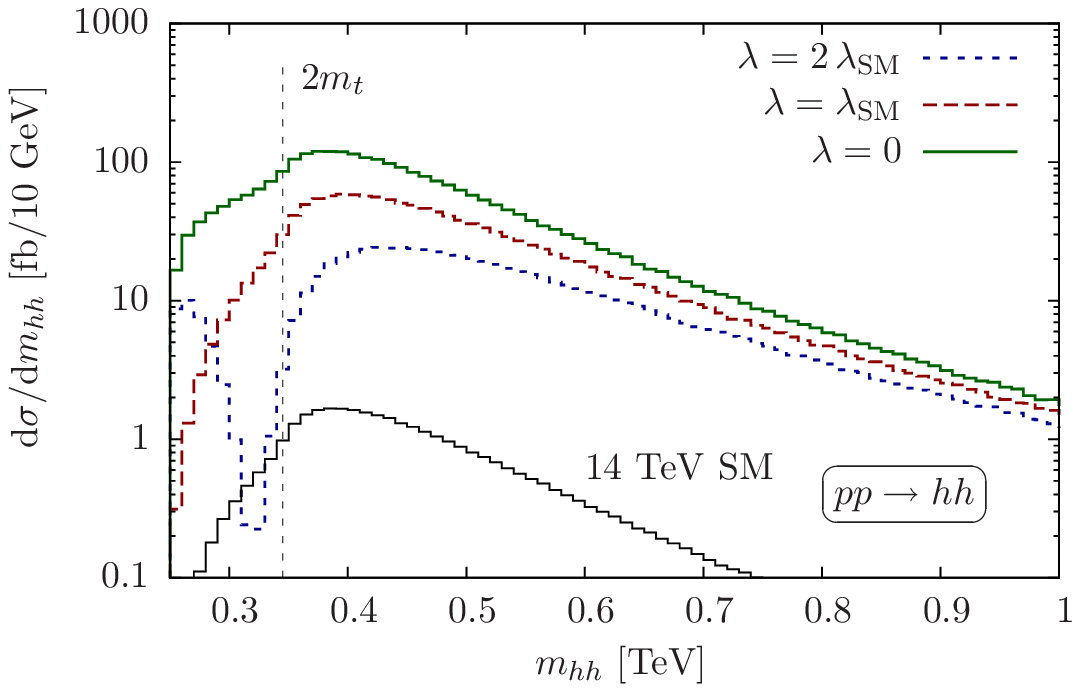}}
  \hfill
  \subfigure[]{\includegraphics[width=0.47\textwidth]{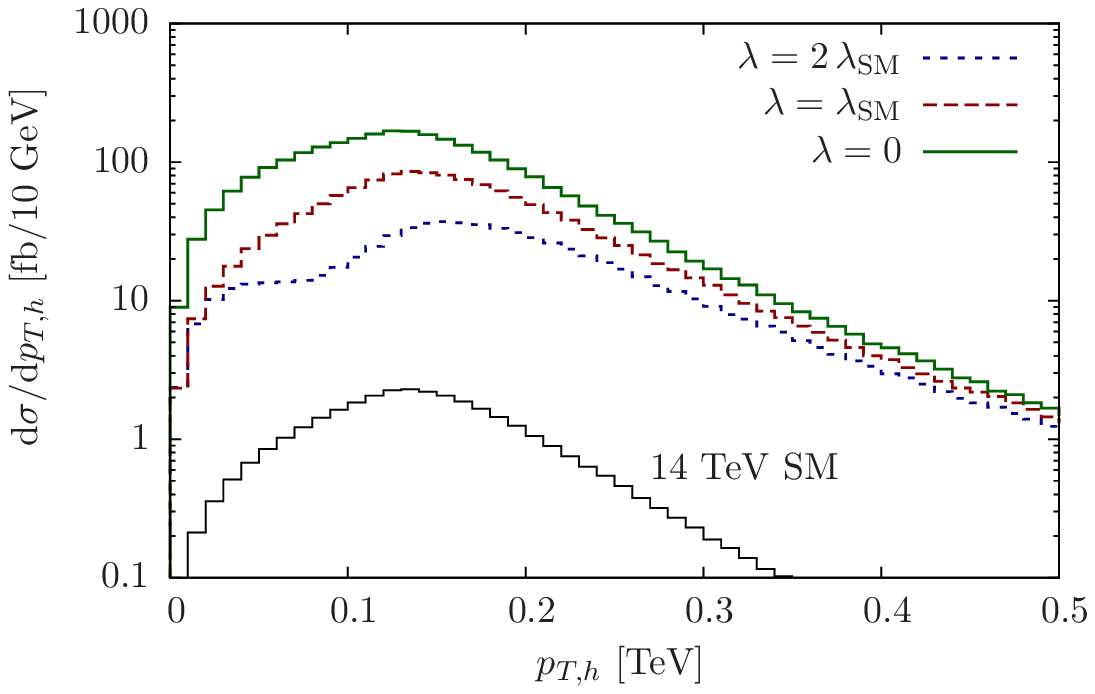}}
  \caption{\label{fig:HHkin} Leading-order parton level distributions
    (including flat NLO normalisation $K$ factors) of the dihiggs
    invariant mass $m_{hh}$ and transverse momentum $p_{T,h}$ for $pp
    \to h h$ at $\sqrt{s} = 100$ TeV for $\lambda=0, \lambda_{\text{SM}}$ and $2\lambda_{\text{SM}}$, shown with the
    $\lambda/\lambda_{\text{SM}}=1$ case for $\sqrt{s}=14$~TeV for
    comparison. }
\end{figure*}
%%%%%%%%%%%%%%%%%%%%%%%%%%%%%%%%%%%
%%%%%%%%%%%%%%%%%%%%%%%%%%%%%%%%%%%
\begin{figure*}[!t]
  \centering
  \subfigure[]{\includegraphics[width=0.47\textwidth]{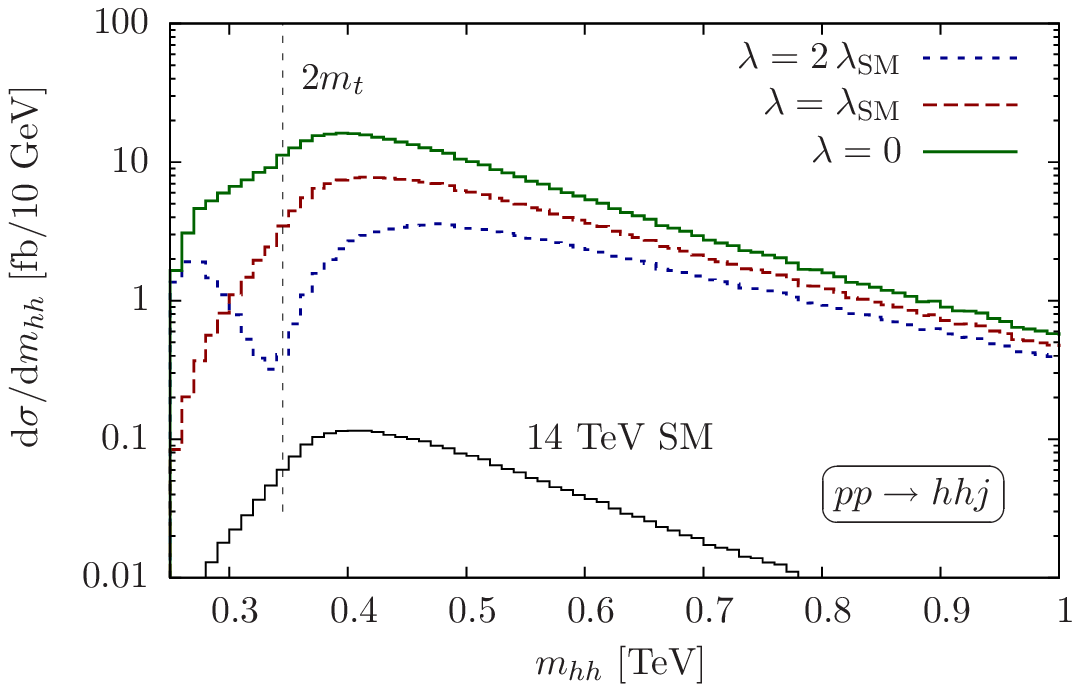}}
  \hfill
  \subfigure[]{\includegraphics[width=0.47\textwidth]{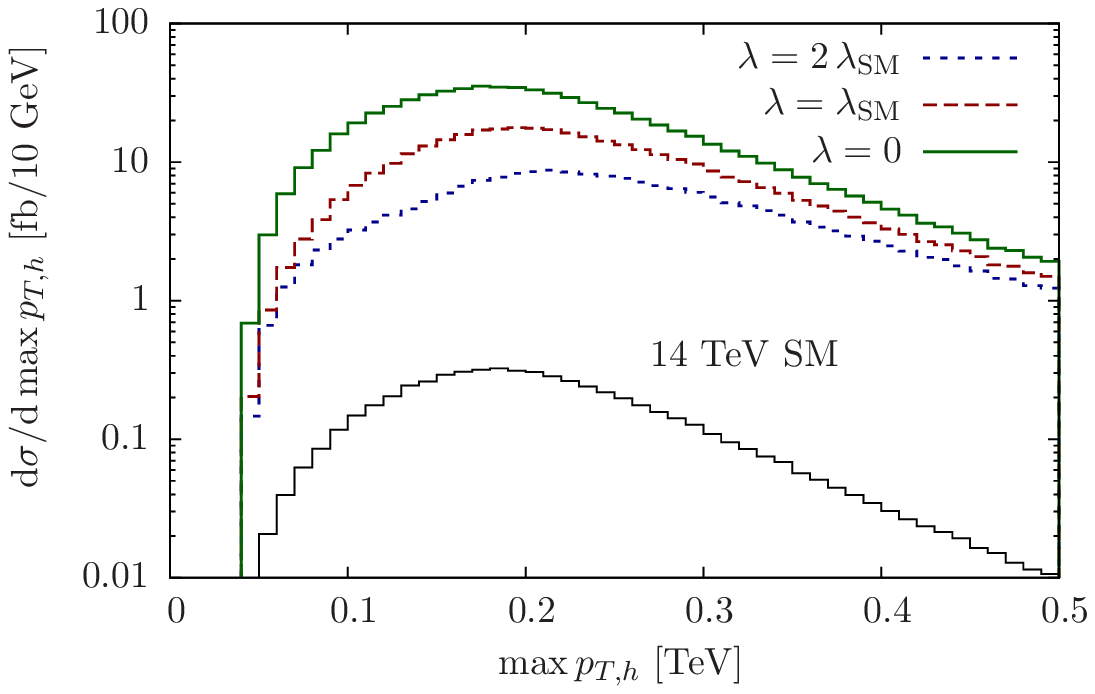}}
  \caption{\label{fig:HHjkin} Leading-order parton level distributions
    of the dihiggs invariant mass $m_{hh}$ and maximum transverse
    momentum $\max p_{T,h}$ for $pp \to h hj$ at $\sqrt{s} = 100$ TeV
    for $p_{T,j}\geq 80~\text{GeV}$ and $|\eta_j|\leq 4.5$,  for $\lambda=0, \lambda_{\text{SM}}$ and $2\lambda_{\text{SM}}$. We also
    include the $\lambda/\lambda_{\text{SM}}=1$ case for
    $\sqrt{s}=14$~TeV for comparison.}
\end{figure*}
%%%%%%%%%%%%%%%%%%%%%%%%%%%%%%%%%%%

Results from other channels suggest a measurement of the Higgs
trilinear at the level of 30-50\% may be possible at the
LHC~\cite{Goertz:2013kp} using a combination of the above channels and
ratios of cross-sections.  The proposed International Linear Collider
could improve on such a measurement if operated with a centre of mass
energy of 1~TeV, in which case an estimated ultimate precision of 13\%
could be achieved~\cite{Dawson:2013bba,Durig:2014lfa}. However,
identifying the possible deviations in Higgs self-couplings due to BSM
physics may require a measurement at greater than even this
accuracy~\cite{Gupta:2013zza,Dawson:2013bba}.

The discovery of new physics and a complete understanding of
electroweak symmetry breaking may therefore require a new high energy
hadron collider~\cite{bsm100}.  A study of the ability of such a
collider to constrain the Higgs trilinear couplings was undertaken as
part of the Snowmass process~\cite{Yao:2013ika,Dawson:2013bba}. While
this study focussed on the $b\bar b \gamma \gamma$ channel, it did not
include any of the dominant backgrounds due to fakes.

In this article we therefore comprehensively analyse the process
\begin{equation}
  p p \to hh +X  \to \left( b + \bar b \right) + (\gamma+\gamma) +X
\label{eq:hhprod}
\end{equation}
at $\sqrt{s}=100$~TeV in order to provide a reliable estimate of the
sensitivity which a very high energy hadron collider would have to
variations in the trilinear Higgs coupling. We also consider the
related same process accompanied by a high transverse momentum jet,
which, as argued in~\cite{us}, accesses new regions of phase space as
well as offering a powerful means to further suppress background
processes at the LHC.

We find that previous studies have substantially overestimated the
performance of a 100~TeV proton-proton collider to measure the Higgs
trilinear coupling. For a 3/ab data sample, we find a sensitivity to
the trilinear coupling of order 30\%, which is comparable to a
measurement at the ILC. For a data set of 30/ab we find an
${\cal{O}}(10\%)$ sensitivity subject to the details of background
systematics.

This work is organised as follows: In Section~\ref{sec:kinematics} we
review the kinematic Higgs distributions at 100~TeV, before presenting
details of our analysis and simulations in Sec.~\ref{sec:analysis}. In
particular, we discuss $hh\to b\bar b \gamma\gamma$ production in
Sec.~\ref{sec:analysis1}, and investigate $hh+{\text{jet}}$ in
Sec.~\ref{sec:analysis2}. We present a combination of the results of
these channels in Sec.~\ref{sec:results}, before we conclude with a
brief discussion and comments on future studies in
Sec.~\ref{sec:conc}.

%%%%%%%%%%%%%%%%%%%%%%%%%%%%%%%%%%%%%%%%%%
\begin{figure*}[!t]
  \centering
  \subfigure[]{\includegraphics[width=0.48\textwidth]{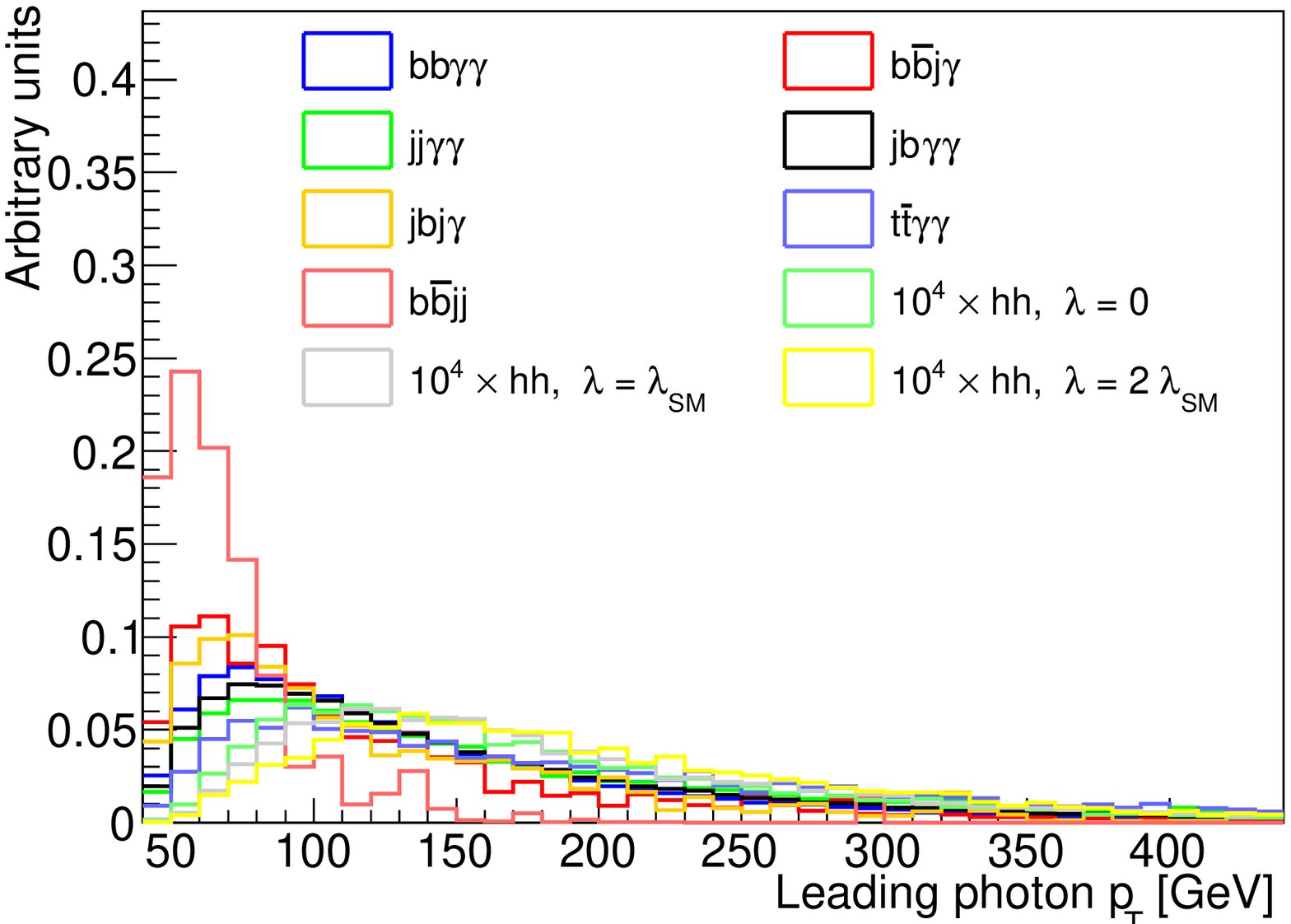}}
  \hfill
  \subfigure[]{\includegraphics[width=0.48\textwidth]{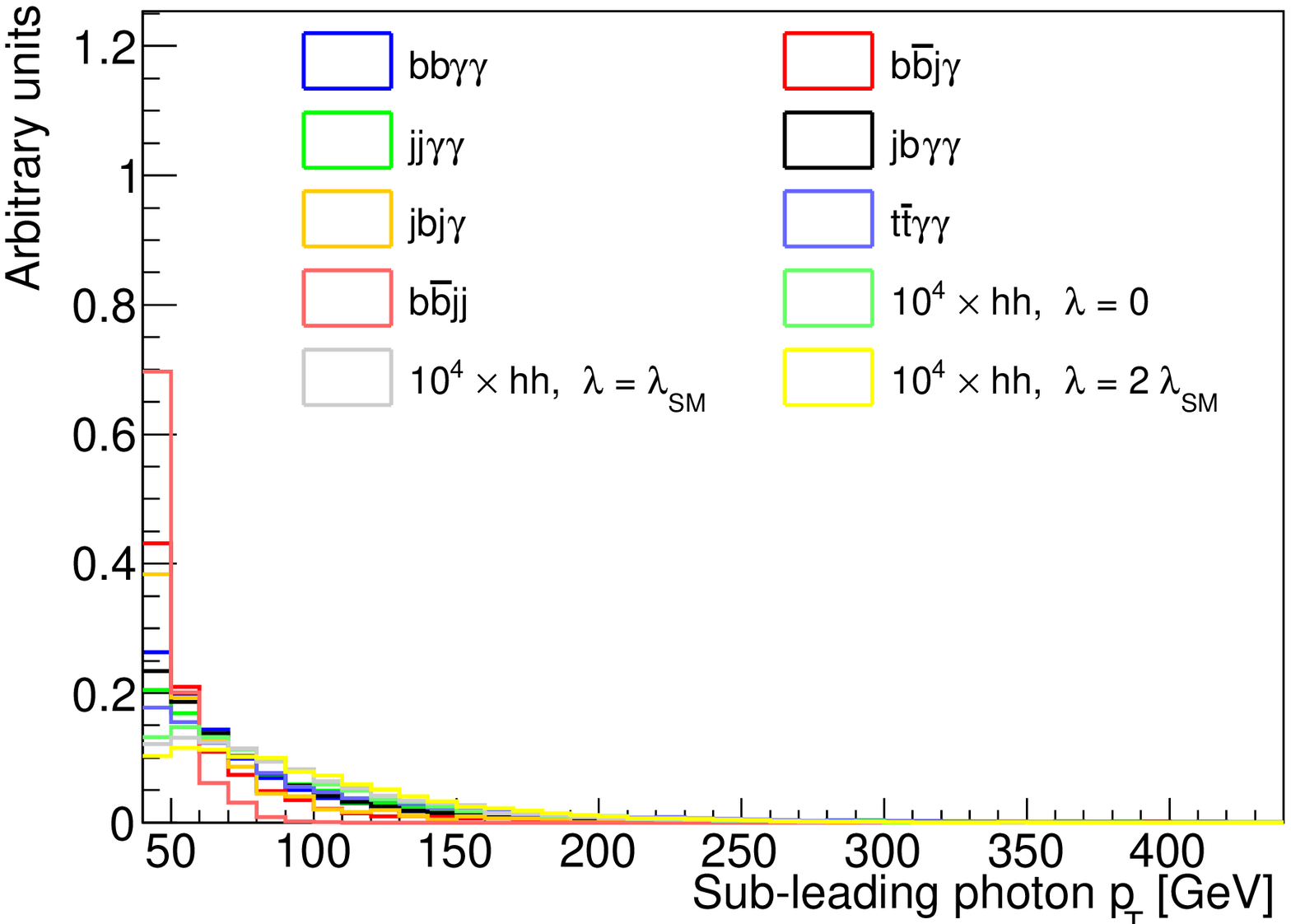}}
  \caption{\label{fig:photon_pt} The left panel (a) shows the
     transverse momentum of the leading photon in $hh\to b\bar  b \gamma \gamma$ events for $\lambda=0,\lambda_{\text{DM}}$ and $2\lambda_{\text{SM}}$ along with various background contributions, while the right panel (b) shows the subleading photon transverse momentum.}
  \vspace{-0.5cm}
\end{figure*}
%%%%%%%%%%%%%%%%%%%%%%%%%%%%%%%%%%%%%%%%%%
%%%%%%%%%%%%%%%%%%%%%%%%%%%%%%%%%%%%%%%%%%
\begin{figure*}[!t]
  \centering
   \subfigure[]{\includegraphics[width=0.48\textwidth]{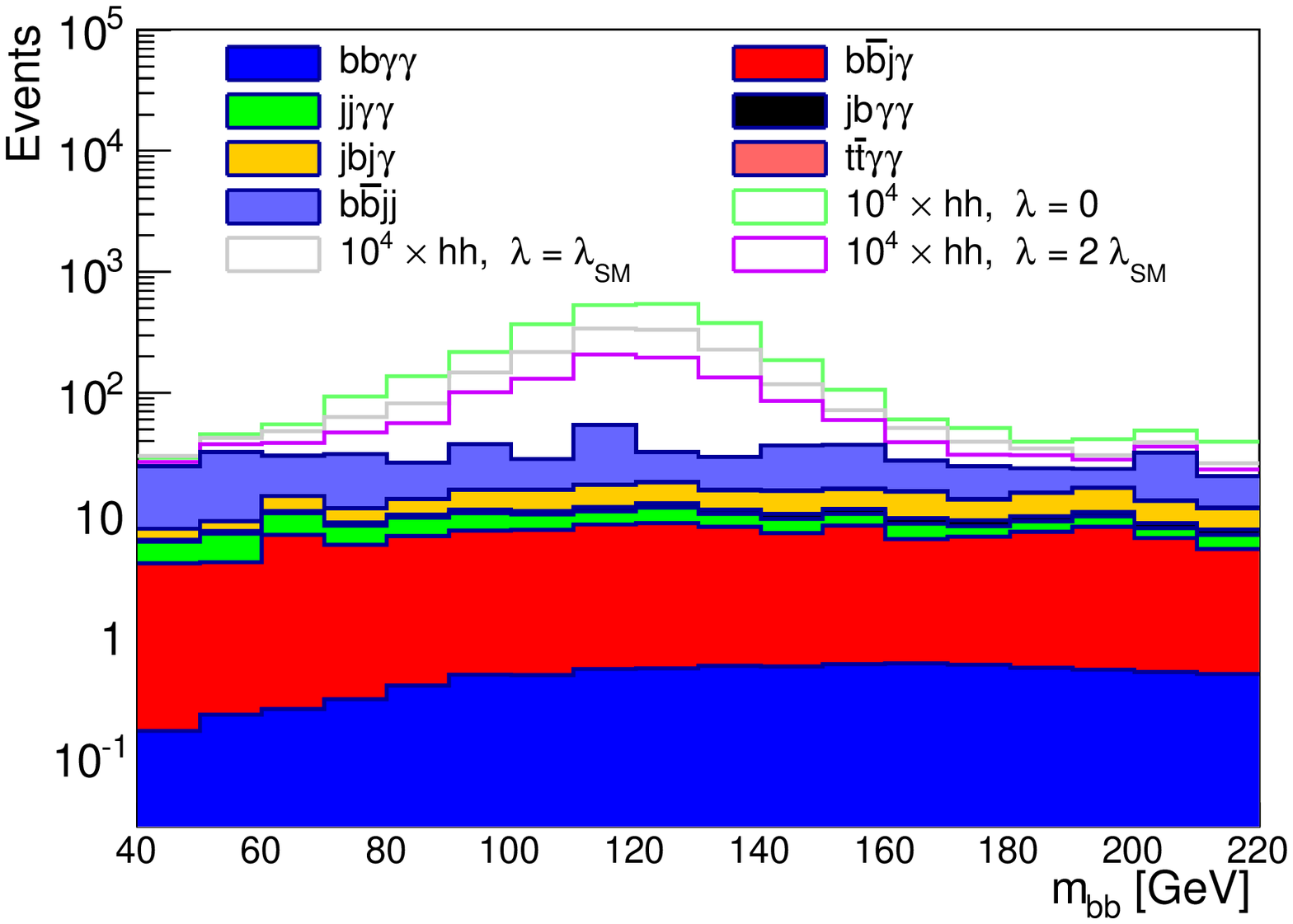}}
  \hfill
  \subfigure[]{\includegraphics[width=0.48\textwidth]{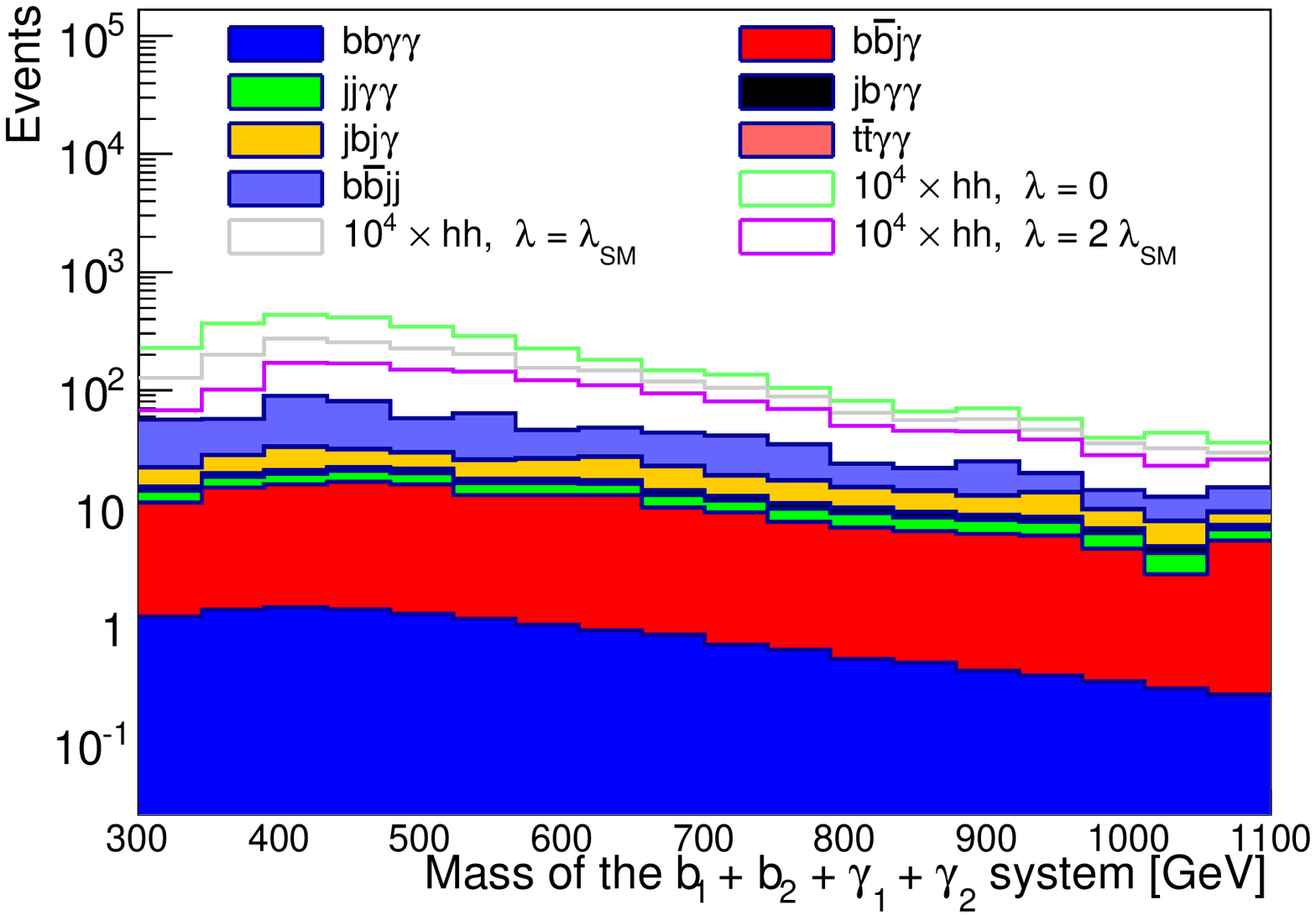}}
  \caption{\label{fig:HH} The left panel  (a) displays the differential
    $m_{b\bar{b}}$ 
    distribution for $\lambda=0,\lambda_{\text{DM}}$ and $2\lambda_{\text{SM}}$ and background contributions. The right  panel (b) shows the invariant mass of
    the 2-photon and 2-$b$-jet system $m_{b\bar{b} \gamma \gamma}$.
  }
  \vspace{-0.5cm}
\end{figure*}
%%%%%%%%%%%%%%%%%%%%%%%%%%%%%%%%%%%%%%%%%%

%%%%%%%%%%%%%%%%%%%%%%%%%%%%%%%%%%%
\section{Kinematics}
\label{sec:kinematics}
%%%%%%%%%%%%%%%%%%%%%%%%%%%%%%%%%%%
We generate signal events at leading order in the Les Houches Event
File format~\cite{Boos:2001cv} using a combination of the
\textsc{Vbfnlo}~\cite{vbfnlo} and
\textsc{FeynArts/FormCalc/LoopTools}~\cite{hahn} frameworks. We
normalise to the NLO cross section by multiplying a phase-space
independent $K$-factor of 1.65~\cite{Dawson:1998py,spira}. 

Our leading order results for $\lambda=(0,1,2)\,\lambda_{\text{SM}}$
are $\sigma_{\text{sig}}=(1676.9,~860.6,~415.5)$~fb
respectively. These are to be compared with an inclusive cross-section
of $33.8~\text{fb}$ at NLO at 14~TeV for $\lambda=\lambda_{\text{SM}}$
\cite{spira}, an increase by a factor of $\sim 40$. To obtain the
cross section after decays to photons and bottom quarks, we multiply
with the branching ratio ${\text{Br}} (hh \to \bar{b}b \gamma
\gamma)\simeq 0.267 \%$.

In Fig.~\ref{fig:HHkin} we show the dihiggs invariant mass $m_{hh}$
and Higgs $p_{T}$ distributions at 100~TeV  for $\lambda=0, \lambda_{\text{SM}}$ and $2\lambda_{\text{SM}}$, with the 14~TeV case for
$\lambda=\lambda_{\text{SM}}$ shown for comparison. While the 100~TeV
distributions have considerably longer tails at high momentum and
invariant mass, they are broadly similar to the 14~TeV ones. In
particular, the peak in the $m_{hh}$ spectrum at around 400~GeV
and the peak in the partonic Higgs transverse momentum just
near $m_t$ due to the diHiggs system being produced near threshold. Due to the interference between the triangle and box
diagrams the region around $s\sim 4m_t^2 $ is most sensitive to
$\lambda$.

This relatively small invariant mass window which provides the most
sensitive probe of $\lambda$ asks for a selection as inclusive as
possible. Such a selection is not possible in the $b\bar b\tau \tau$
and $b\bar b W^+W^-$ modes, as they crucially rely on the boosted
kinematics regime. However, as demonstrated in \cite{us}, lower
invariant dihiggs masses can be obtained by recoiling the dihiggs
system against a hard jet. Such a process becomes increasingly likely
when we increase the centre of mass energy as energetic jet radiation
becomes unsuppressed. Indeed, as displayed in Fig.~\ref{fig:HHjkin},
the region of sensitivity to $\lambda$ is reduced for recoils at
$p_{T,j}\geq 80~\text{GeV}$. However, the price to be paid is in
smaller total cross sections which we compute at leading order to be
$(494.5,262.9,149.3)$~fb for $\lambda=(0,1,2)\,\lambda_{\text{SM}}$
for jets with $|\eta_j|<4.5$.

The Higgs bosons in very high energy dihiggs events are typically
produced in the central pseudorapidity region. For the inclusive $hh$
case it is important to stress that a considerable fraction of the
cross section stems from relatively small scattering angles at large
pseudorapidity. Hence it is desirable to have as much forward detector
coverage as possible to access these events at a 100 TeV collider.

%%%%%%%%%%%%%%%%%%%%%%%%%%%%%%%%%%%%%%%%%%
\begin{table*}[!t]
\begin{center}
%\begin{sidewaystable}
\resizebox{\linewidth}{!}{
\begin{tabular}{|c|c|c|c|c|c|c|c|}
\hline
 Sample & $\Delta R(b_1,b_2)$ & $\Delta R(\gamma_1,\gamma_2)$ & $p_{T,\gamma\gamma}$ & $p_{T,bb}$ & $\Delta \phi(b_{2}, \gamma\gamma)$ & $m_{bb}$ & $m_{\gamma\gamma}$ \\
\hline
$(h \rightarrow b\bar{b})(h \rightarrow \gamma \gamma) \lambda = \lambda_{SM}$ &  1.18\tomone & 1.05\tomone & 9.76\tomtwo & 8.40\tomtwo & 6.85\tomtwo & 5.96\tomtwo & 5.96\tomtwo \\ 
$(h \rightarrow b\bar{b})(h \rightarrow \gamma \gamma) \lambda = 0$ &  1.93\tomone & 1.68\tomone & 1.54\tomone & 1.29\tomone & 1.03\tomone & 8.88\tomtwo & 8.87\tomtwo \\ 
$(h \rightarrow b\bar{b})(h \rightarrow \gamma \gamma) \lambda = 2 \lambda_{SM}$ &  6.74\tomtwo & 6.24\tomtwo & 5.95\tomtwo & 5.30\tomtwo & 4.55\tomtwo & 3.91\tomtwo & 3.91\tomtwo \\ 
\hline
$jj\gamma\gamma$               &  2.76\toone & 8.94 & 5.99 & 4.46 & 3.88 & 1.48 & 7.20\tomtwo \\
$bbj\gamma$                        &  5.97\toone & 2.01\toone & 1.08\toone & 8.75 & 8.18 & 3.04 & 1.43\tomtwo\\
$b\bar{b}jj$               & 1.99\totwo & 4.79\toone & 1.47\toone & 7.82 & 7.67 & 2.81\tomone & 5.06\tomthree \\ 
$t\bar{t}\gamma\gamma$ &  1.01 & 4.31\tomone & 3.62\tomone & 2.78\tomone & 2.22\tomone & 9.06\tomtwo & 3.38\tomtwo \\ 
$b\bar{b}\gamma\gamma$               & 2.70 & 8.26\tomone &  5.80\tomone & 4.58\tomone &  4.48\tomone & 1.69\tomone & 1.21\tomtwo \\
$jbj\gamma$ &  3.61\toone & 8.37 & 5.70 & 4.34 & 3.88 & 7.24\tomone & 3.04\tomthree \\ 
$jb\gamma\gamma$               &  5.18 & 1.57 & 9.86\tomone & 7.91\tomone & 6.99\tomone & 2.41\tomone & 8.57\tomthree \\
\hline
Background &  3.31\totwo & 8.81\toone & 3.91\toone & 2.69\toone & 2.50\toone & 6.03 & 1.49\tomone \\ 
\hline
$S/B$ ($\lambda/\lambda_{SM} = 0$) &  0.00058 & 0.0019 & 0.0039 & 0.0048 & 0.0041 & 0.015 & 0.59 \\ 
$S/\sqrt{B}$ ($\lambda/\lambda_{SM} = 0$) &  0.58 & 0.98 & 1.35 & 1.36 & 1.13 & 1.98 & 12.58 \\ 
\hline
$S/B$ ($\lambda/\lambda_{SM} = 1$) &  0.00036 & 0.0012 & 0.0025 & 0.0031 & 0.0027 & 0.0099 & 0.4 \\ 
$S/\sqrt{B}$ ($\lambda/\lambda_{SM} = 1$) &  0.36 & 0.62 & 0.85 & 0.89 & 0.75 & 1.33 & 8.45 \\ 
\hline
$S/B$ ($\lambda/\lambda_{SM} = 2$) &  0.0002 & 0.00071 & 0.0015 & 0.002 & 0.0018 & 0.0065 & 0.26 \\ 
$S/\sqrt{B}$ ($\lambda/\lambda_{SM} = 2$) &  0.20 & 0.36 & 0.52 & 0.56 & 0.50 & 0.87 & 5.54 \\  \hline
\end{tabular}
}
\label{tab:hh}
%\end{sidewaystable}
\end{center}
\caption{This table shows the cutflow and cross-sections for the $b\bar b \gamma\gamma$ analysis. The cross sections are given in femtobarns, and $S/\sqrt{B}$ 
  is shown for a luminosity of 3000/fb.
  After the pre-selection described  on the text, the $\Delta R$ between the two leading $b$-jets
  and the two leading photons is required to be $<1.7$,  and the cross section after this selection is shown
  in the second and third columns.
  The transverse momentum of the 2-photons and 2-$b$-jets systems are required to be $> 150$ GeV, as shown in the fourth
  and fifth columns.
  The $\Delta \phi$ between the sub-leading $b$-jet and the diphoton system is required to be  $\Delta \phi > 1.6$.
  Finally the cuts on the invariant mass of the two hardest $b$-jets and photons are implemented as $|m_{bb} - 120\textrm{ GeV}| < 30$ GeV and
  $|m_{\gamma\gamma} - 125\textrm{ GeV}| < 1$ GeV. 
  Further details on the cuts can be found in the text.
  }
\end{table*}
%%%%%%%%%%%%%%%%%%%%%%%%%%%%%%%%%%%%%%%%%%

%%%%%%%%%%%%%%%%%%%%%%%%%%%%%%%%%%%%%%%
\section{Analysis}
\label{sec:analysis}
%%%%%%%%%%%%%%%%%%%%%%%%%%%%%%%%%%%%%%

%%%%%%%%%%%%%%%%%%%%%%%%%%%%%%%%%%%%%%
\subsection*{Event Generation and Detector Simulation}
%%%%%%%%%%%%%%%%%%%%%%%%%%%%%%%%%%%%%%
We generate the QCD and electroweak background events using
\textsc{MadGraph 5}~\cite{Alwall:2011uj}, which are showered and
hadronised with \textsc{Pythia 8}~\cite{Sjostrand:2007gs}. Of
particular importance in this channel are the so-called reducible
backgrounds where jets can fake a hard photon. For all the backgrounds
we use the leading order cross sections as obtained from
{\sc{MadEvent}}. In our analysis for $hh \to \bar{b}b \gamma \gamma$
we consider all reducible and irreducible backgrounds with at least four
reconstructed objects in the final state without merging.

Because the irreducible and reducible backgrounds for this final state
are large compared to the signal, we devote particular care to
simulating fake rates. However, we stress that our parametrisation of
fake rates and the detector response is based on the present
performance of ATLAS and CMS and will likely deviate from that of an
envisioned detector designed for $\sqrt{s}=100$ TeV.

When reconstructing the final state objects we consider all visible
particles with $|\eta| < 5.0$. We smear the momenta of all
reconstructed final state objects with Gaussians, using the
parametrisations of~\cite{param1} for jets and muons, as well
as a 95\% jet reconstruction efficiency, and we take the electron
smearing parametrisation from~\cite{param2}.  The photons
were smeared using a Gaussian with standard deviation of $0.1$\% of the photon $p_T$.
We simulate $b$-tagging
by matching a jet with a hadron containing a bottom or charm before
decay and multiply a flat $b$-tagging efficiency of 70\%, a mis-tag
rate of 10\% for $c$-jets and 1\% for light-flavor jets. We assume the
jet-faking-lepton and jet-faking-photon probabilities to be
momentum-dependent and parametrise them to be $\mathcal{P}_{j\to l} =
0.0048\times e^{-0.035 p_{T,j}/\text{GeV}}$ and $\mathcal{P}_{j\to \gamma}
=0.0093\times e^{-0.036 p_{T,j}/\text{GeV}}$, respectively. We further take
into account jet, photon and muon detection efficiencies parametrised
according to \cite{param1}, while the electron efficiency is taken
from \cite{param2}.  We do not distinguish between the tagging rates
in the barrel and endcaps.

The detector parametrisation used is very conservative, particularly
on the photon identification efficiency, which is parametrised as
$\mathcal{E}_{\gamma} = 0.76 - 1.98
e^{-p_{T,\gamma}/16.1~\text{GeV}}$.  This efficiency performance has a
turn on curve that only reaches a $> 70\%$ efficiency at $\sim 60$
GeV, while for a $20$ GeV photon, its detection efficiency is only
$18\%$. This is a significant limitation on the analysis, as the
photons from the Higgs decay are expected to have often a lower $p_T$,
as can be seen in Fig.~\ref{fig:photon_pt}.
The results can be improved if one is allowed to reduce the photon
$b$-jet transverse momentum thresholds in 100 TeV machine, with turn on
curve reaching a stable efficiency at a lower transverse momentum.
This would
increase the signal acceptance in the analysis, and it would
open space for more elaborate techniques for background rejection.

%%%%%%%%%%%%%%%%%%%%%%%%%%%%%%%%%%%%%%%%%%
\begin{figure*}[!t]
  \centering
  \subfigure[]{\includegraphics[width=0.48\textwidth]{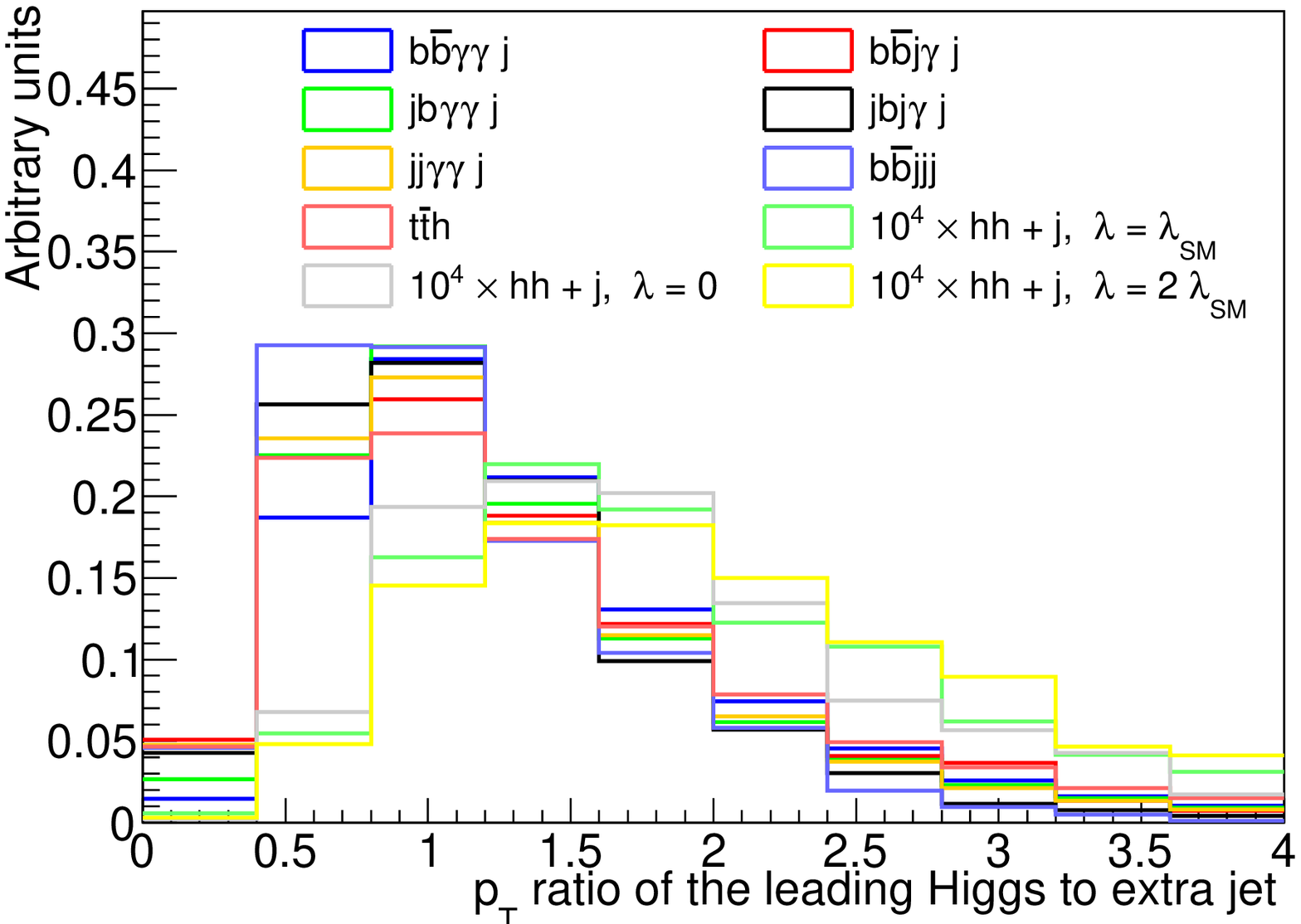}}
  \hfill
  \subfigure[]{\includegraphics[width=0.48\textwidth]{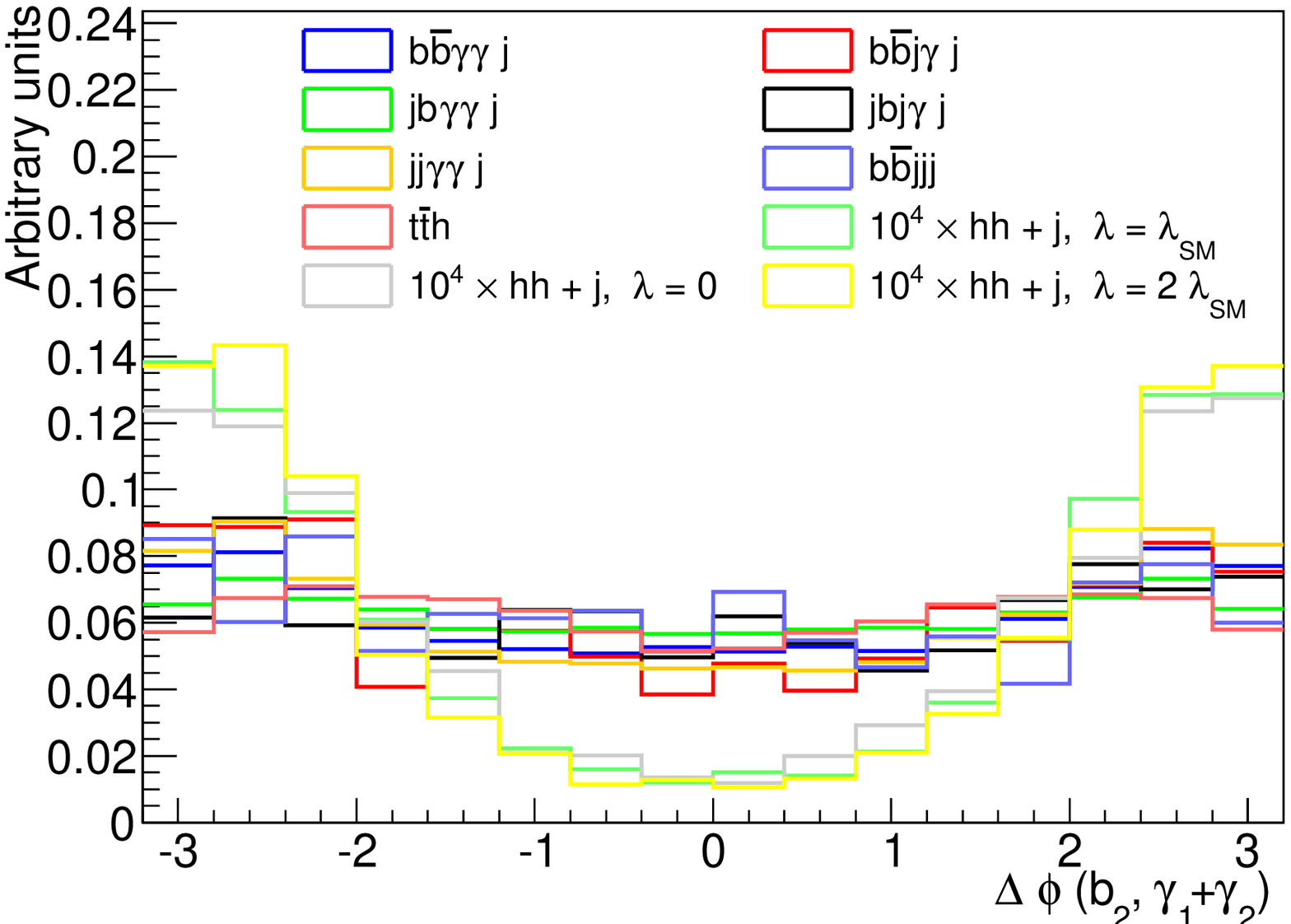}}
  \caption{\label{fig:HHj_cuts} The left panel (a) shows the ratio of
    the transverse momentum of  leading reconstructed Higgs to the
    transverse momentum of the extra jet for $\lambda=0,\lambda_{\text{SM}}$ and $2\lambda_{\text{SM}}$ as well as the backgrounds. The right panel (b) shows
    the $\Delta \phi$ between the subleading $b$-let and the
    $\gamma\gamma$ system for the same data.}
\end{figure*}
%%%%%%%%%%%%%%%%%%%%%%%%%%%%%%%%%%%%%%%%%%
%%%%%%%%%%%%%%%%%%%%%%%%%%%%%%%%%%%%%%%%%%%%%%%%%%%%%%%%
\begin{table*}[!t]
%\begin{sidewaystable}
\begin{center}
\resizebox{\linewidth}{!}{
\begin{tabular}{|c|c|c|c|c|c|c|c|c|c|}
\hline
 Sample & Pre-selected & $N_{jets} \ge 3$ & Extra jet $p_T$ & $p_{T,\gamma\gamma}$ & $m_{bb}$ & $m_{\gamma\gamma}$ & $N_{\textrm{jets}} \le 6$ & $\Delta \phi(b_{2}, \gamma\gamma)$ & $p_{T,h1}/p_{T,j}$ \\
\hline
$hh(b\bar{b} \gamma \gamma) + j, \lambda = 0$ & 5.85\tomtwo & 4.98\tomtwo & 3.57\tomtwo & 3.10\tomtwo & 2.23\tomtwo & 2.22\tomtwo & 2.16\tomtwo & 1.76\tomtwo & 1.61\tomtwo \\ 
$hh(b\bar{b} \gamma \gamma) + j, \lambda/\lambda_{SM} = 1$ & 3.56\tomtwo & 3.08\tomtwo & 2.18\tomtwo & 1.91\tomtwo & 1.36\tomtwo & 1.35\tomtwo & 1.32\tomtwo & 1.13\tomtwo & 1.06\tomtwo \\ 
$hh(b\bar{b} \gamma \gamma) + j, \lambda/\lambda_{SM} = 2$ & 2.09\tomtwo & 1.85\tomtwo & 1.33\tomtwo & 1.21\tomtwo & 8.50\tomthree & 8.49\tomthree & 8.30\tomthree & 7.20\tomthree & 6.91\tomthree \\ 
\hline
$b\bar{b}\gamma\gamma j$ (QED=2) & 7.62 & 6.74 & 5.21 & 2.83 & 4.39\tomone & 2.48\tomthree & 2.48\tomthree & 2.04\tomthree & 1.75\tomthree \\ 
$b\bar{b}\gamma\gamma j$ (QED=4) & 1.39\tomone & 1.25\tomone & 9.80\tomtwo & 6.05\tomtwo & 2.56\tomtwo & 9.32\tomfive & 9.32\tomfive & 7.96\tomfive & 7.96\tomfive \\ 
\hline
$t\bar{t}h(\rightarrow \gamma\gamma)$ & 7.34\tomone & 6.91\tomone & 5.32\tomone & 4.02\tomone & 4.44\tomtwo & 4.39\tomtwo & 3.76\tomtwo & 2.55\tomtwo & 2.03\tomtwo \\ 
\hline
$jj\gamma\gamma j$ & 1.66\toone & 1.55\toone & 1.26\toone & 7.87 & 1.46 & 7.10\tomthree & 6.95\tomthree & 4.58\tomthree & 4.41\tomthree \\ 
$b\bar{b}j\gamma j$ & 4.84\toone & 4.48\toone & 3.55\toone & 1.88\toone & 3.25 & 8.51\tomtwo & 8.49\tomtwo & 1.46\tomthree & 1.17\tomthree \\ 
$jb\gamma\gamma j$ & 5.20 & 4.76 & 3.75 & 2.15 & 2.02\tomone & 1.52\tomthree & 1.52\tomthree & 8.67\tomfour & 6.88\tomfour \\ 
$jbj\gamma j$ & 1.65\toone & 1.57\toone & 1.29\toone & 6.55 & 4.58\tomone & 1.25\tomthree & 1.24\tomthree & 8.75\tomfour & 5.97\tomfour \\ 
$b\bar{b}jjj$ & 3.10\toone & 2.93\toone & 2.23\toone & 4.52 & 7.54\tomone & 1.30\tomthree & 1.06\tomthree & 7.29\tomfour & 2.44\tomfour \\ 
\hline
Background & 1.26\totwo & 1.18\totwo & 9.28\toone & 4.32\toone & 6.64 & 1.43\tomone & 1.36\tomone & 3.61\tomtwo & 2.93\tomtwo \\ 
\hline
$S/B$ ($\lambda/\lambda_{SM} = 0$) & 0.00046 & 0.00042 & 0.00038 & 0.00072 & 0.0034 & 0.16 & 0.16 & 0.49 & 0.55 \\ 
$S/\sqrt{B}$ ($\lambda/\lambda_{SM} = 0$) & 0.29 & 0.25 & 0.20 & 0.26 & 0.47 & 3.22 & 3.21 & 5.08 & 5.17 \\ 
\hline
$S/B$ ($\lambda/\lambda_{SM} = 1$) & 0.00028 & 0.00026 & 0.00023 & 0.00044 & 0.002 & 0.095 & 0.097 & 0.31 & 0.36 \\ 
$S/\sqrt{B}$ ($\lambda/\lambda_{SM} = 1$) & 0.17 & 0.16 & 0.12 & 0.16 & 0.29 & 1.96 & 1.96 & 3.26 & 3.39 \\ 
\hline
$S/B$ ($\lambda/\lambda_{SM} = 2$) & 0.00017 & 0.00016 & 0.00014 & 0.00028 & 0.0013 & 0.059 & 0.061 & 0.2 & 0.24 \\ 
$S/\sqrt{B}$ ($\lambda/\lambda_{SM} = 2$) & 0.10 & 0.09 & 0.08 & 0.10 & 0.18 & 1.23 & 1.23 & 2.08 & 2.21 \\ 
\hline
\end{tabular}
}
\end{center}
%\end{sidewaystable}
\caption{This table shows the cutflow for the $hh+j$ analysis. The cross sections are given in femtobarn, and $S/\sqrt{B}$ is 
  shown for a luminosity of 3000/fb.
  After the pre-selection described in the text, a jet multiplicity requirement is implemented
  to guarantee there is one extra jet besides the two Higgs $b$-jets and the cross section after this
  requirement is shown in the third column.
  The transverse momentum of the extra jet is required to be greater than $100$ GeV in the following column.
  The transverse momentum of the two hardest photons is required to be greater than $160$ GeV in the fifth column, and is followed
  by the Higgs mass requirements of $|m_{bb} - 120\textrm{ GeV}| < 30$ GeV and $|m_{\gamma\gamma} - 125\textrm{ GeV}| < 1$ GeV.
  The $t\bar{t}h$ background is reduced by the jet multiplicity requirement that there are $\leq 6$ jets in the eighth column.
  The $\Delta \phi$ selection between the subleading $b$-jet and the hardest two photons system is required to be greater than $1.6$ in the next column.
  Finally the transverse momentum ratio between the leading reconstructed Higgs and the extra jet is required to be greater than 1.
  Further details on the cuts can 
  be found in the text.
  }
\vspace{-0.5cm}
\end{table*}
%%%%%%%%%%%%%%%%%%%%%%%%%%%%%%%%%%%%%%%%%%
%%%%%%%%%%%%%%%%%%%%%%%%%%%%%%%%%%%%%%%%%%
\begin{figure*}[!t]
  \centering
  \subfigure[]{\includegraphics[width=0.48\textwidth]{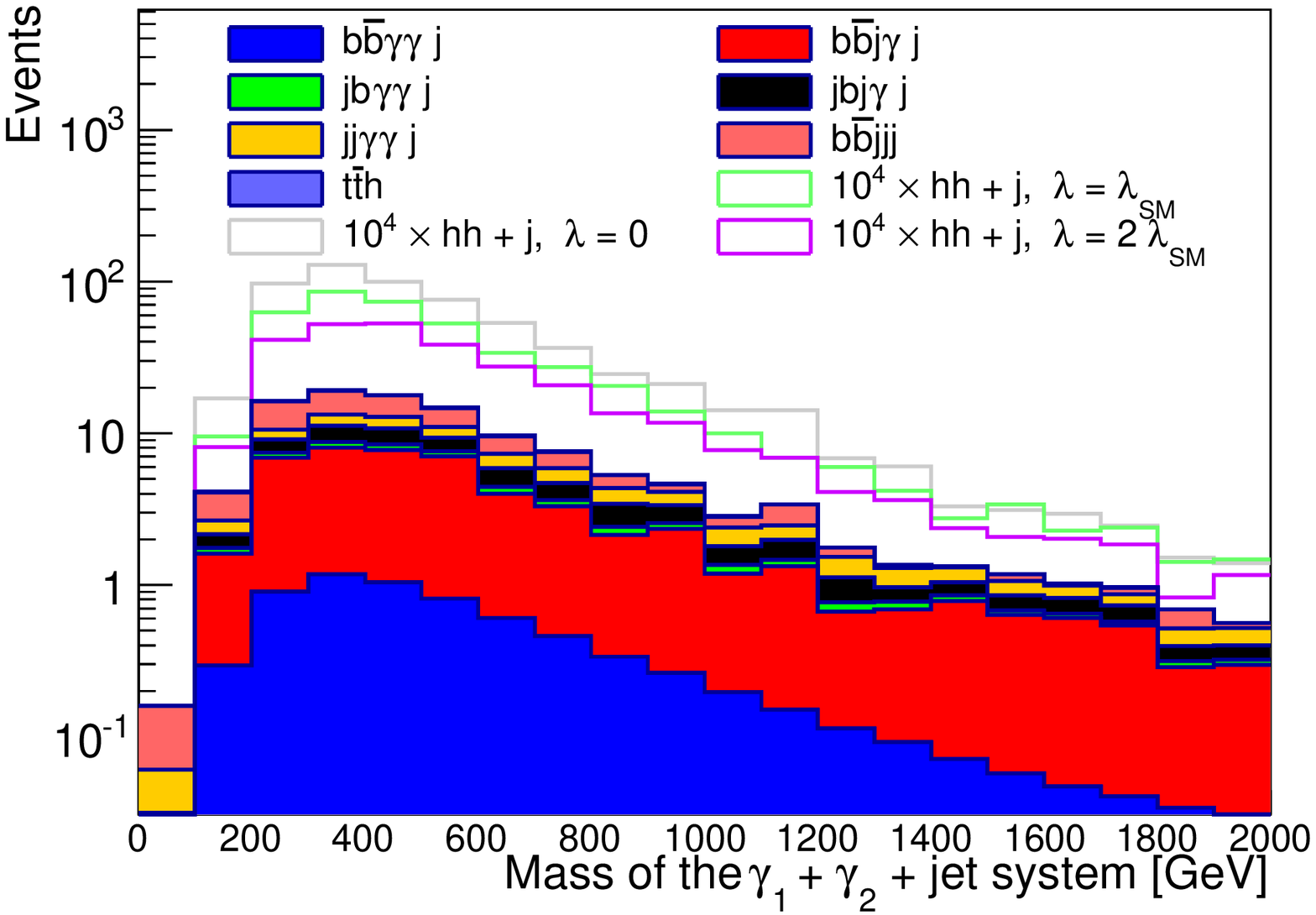}}
  \hfill
  \subfigure[]{\includegraphics[width=0.48\textwidth]{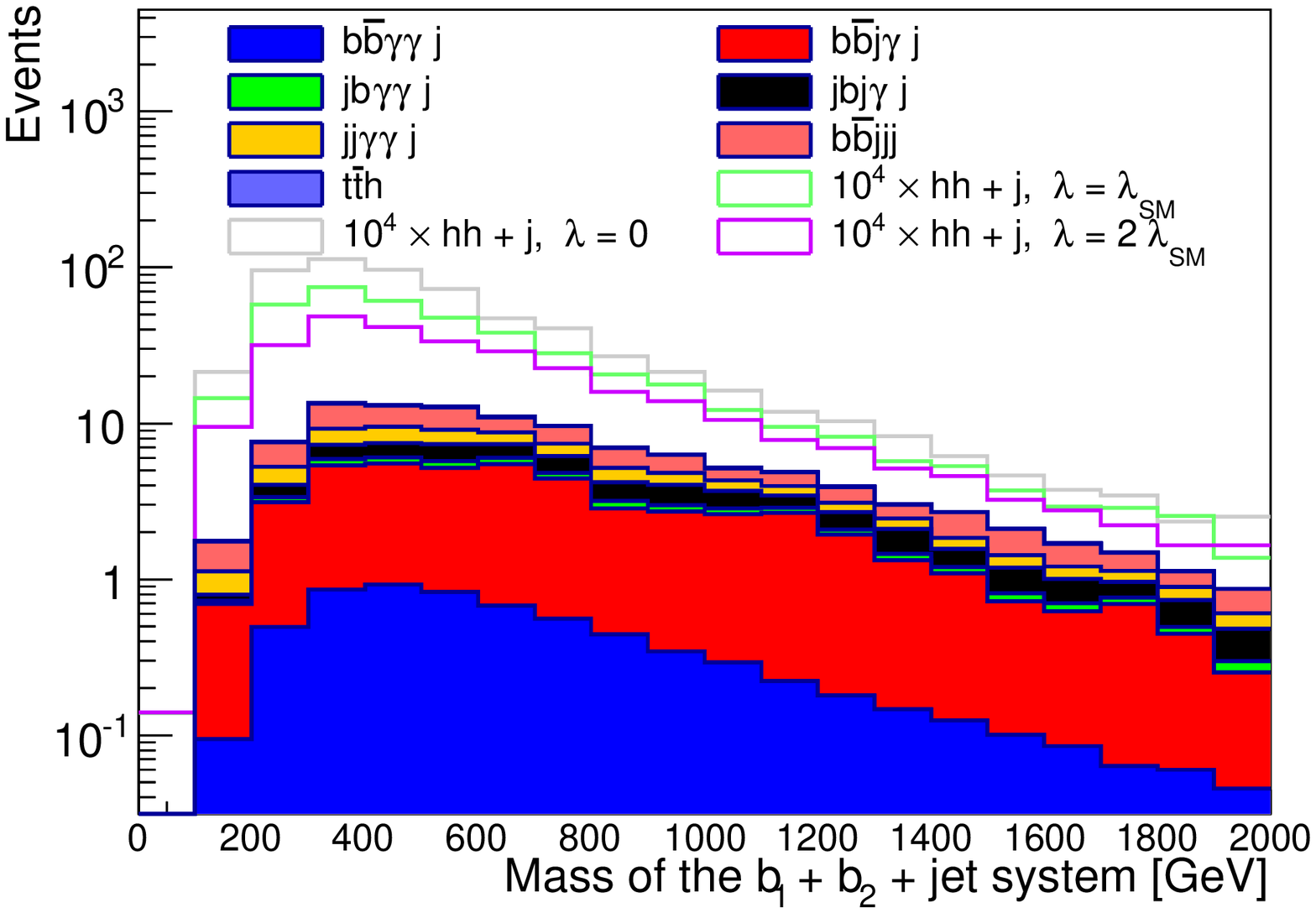}}\\
  \vspace{-0.3cm}
  \subfigure[]{\includegraphics[width=0.48\textwidth]{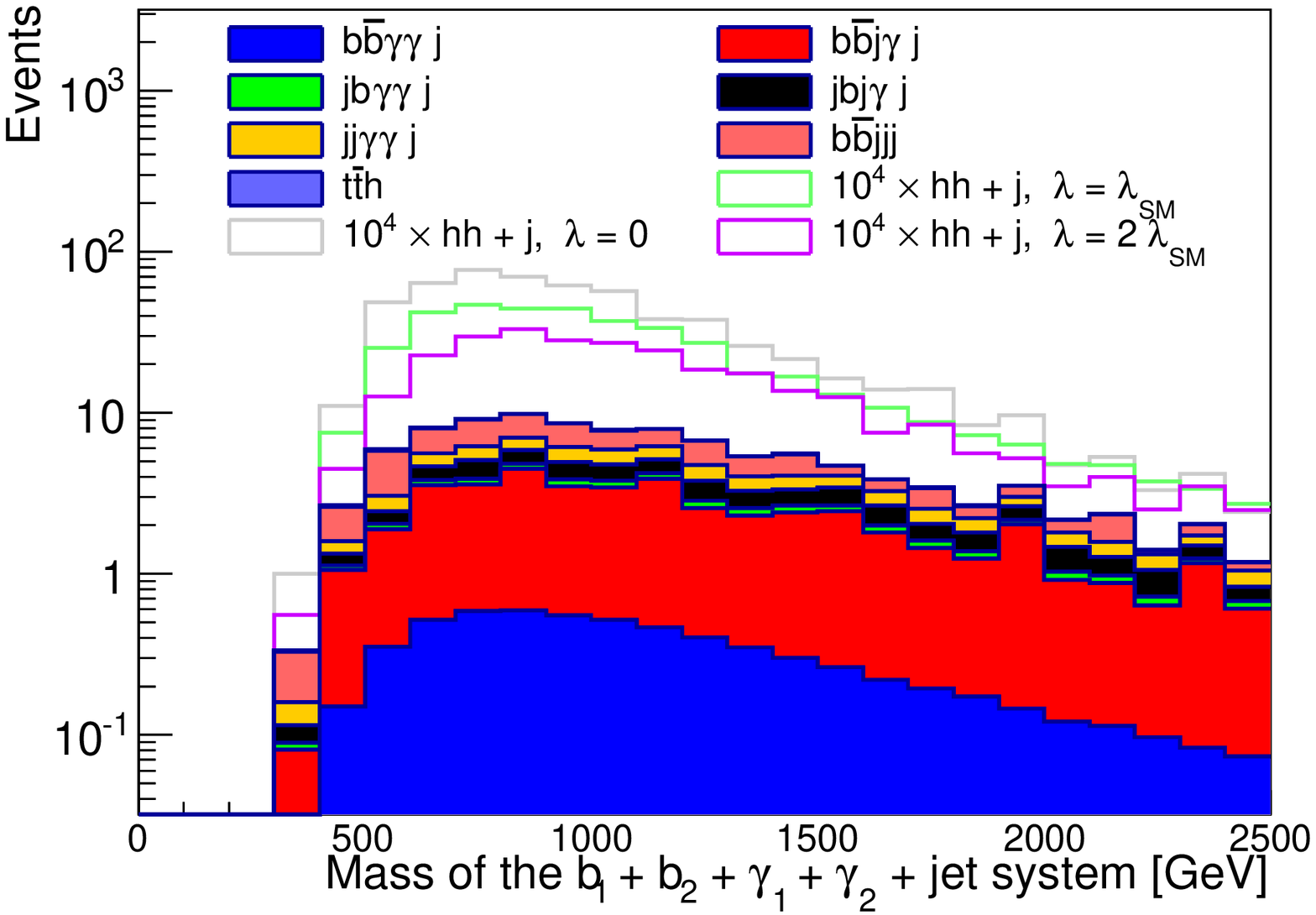}}
  \hfill
  \parbox{0.48\textwidth}{ \vspace{-5cm}\caption{\label{fig:HHj} Panel
      (a) shows the invariant mass distribution of the two hardest
      isolated photons and the extra jet  $m_{\gamma
        \gamma j}$ for the $hh+\text{jet}$ analysis. Panel (b) displays $m_{b\bar{b}j}$ and panel (c)
      shows the invariant mass of the 2-photon, 2-$b$-jet and extra jet
      system $m_{b\bar{b} \gamma \gamma j}$. We show the signal
      distributions for $\lambda=0, \lambda_{\text{SM}}$ and
      $2\lambda_{\text{SM}}$ and the backgrounds in all cases.}}
\end{figure*}
%%%%%%%%%%%%%%%%%%%%%%%%%%%%%%%%%%%%%%%%%%

%%%%%%%%%%%%%%%%%%%%%%%%%%%%%%%%%%%%%%%%%%%%%%%%%%%%%%5
\subsection{$hh \to b \bar{b} \gamma \gamma$}
\label{sec:analysis1}
%%%%%%%%%%%%%%%%%%%%%%%%%%%%%%%%%%%%%%%%%%%%%%%%%%%%%%%%
To reconstruct the $b\bar{b} \gamma \gamma$ final state we require two
reconstructed anti-$k_T$ jets with $R=0.4$ and $p_T>40$ GeV within
$|\eta| < 3.0$. The jets are recombined using
{\sc{FastJet}}~\cite{Cacciari:2011ma}. For the photons we require
$p_T>40$ GeV and $|\eta| < 3.0$. To ensure the photons are isolated we
sum the energy of the visible particles in a cone of $R=0.3$ around
the photon and we only accept them if
$p_{T,{\text{vis}}}/p_{T,{\gamma}} \leq 0.05$. Likewise, we reject a
jet if $\Delta R_{\mathrm{jet}, \gamma} < 0.3$ for any jet-$\gamma$
combination and for $\Delta R_{\gamma, \gamma} < 0.4$ we reject the
softer photon.  For both the jets and photons we smear the
four-momenta of the reconstructed objects as mentioned in the previous
section. 

To identify isolated leptons with $p_T>40$ GeV we apply the same
isolation requirement as for the photons. To accept an event we
require two $b$-jets and two photons.  Events with one or more
isolated leptons are vetoed. At this stage of the analysis we find a
small signal-over-background ratio of $S/B\simeq 3 \times 10^{-4}$ and
$S/\sqrt{B}\simeq 0.28$ after $3000/{\text{fb}}$.

To enhance $S/B$ we apply cuts on the maximum angular separation and
the vectorial sum of the transverse momentum of the two hardest
photons and $b$-jets respectively (see Tab.~\ref{tab:hh} for a
detailed cut flow). In particular we require $\Delta R_{b_1,b_2} < 1.7$, $\Delta
R_{\gamma_1,\gamma_2}< 1.7$, $p_{T,{b \bar b}} > 150$ GeV and
$p_{T,{\gamma \gamma}} > 150$ GeV. As a next step, a selection requirement
on the $\phi$ difference between the sub-leading $b$-jet and the two hardest photons
is required to be  greater than $1.6$. After applying these kinematic cuts
we find $S/B \simeq 3 \times 10^{-3}$ and $S/\sqrt{B} \simeq 0.9$.

Finally the $b$-jets and photons are recombined to the Higgs mass with
$|m_{b\bar b} -120| < 30$ GeV and $|m_{\gamma \gamma} -125| < 1$
GeV. The narrow window for the invariant mass of the
di-photon system allows the rejection of a large fraction of the backgrounds
and improves the statistical significance of the analysis to $S/B
\simeq 0.4$ and $S/\sqrt{B} \simeq 8.45$ for $\lambda =
\lambda_{\text{SM}}$.

Compared to the ATLAS analysis in \cite{bbaaAtlas}, the transverse momentum
requirement for the photons and $b$-jets is required to be stricter in an attempt
to control the effect of the pile up contribution, while otherwise the pre-selection
is made in a very similar way, including the veto on isolated leptons. While no pile up jets
were added or simulated, the parametrisation used for the objects' reconstruction and
identification include the effect of the pile up in the detector performance.

The parametrisation
used in this study for the detector is similar to the one used in the~\cite{bbaaAtlas}. Two exceptions are the $b$-tagging performance, which was taken as a constant and not depending on the transverse
momentum, and the Gaussian smearing functions for the photons have a $0.1\% \times p_T$ standard deviation in our study.
The $\Delta R$ requirements on the hardest $b$-jets and photons were also taken
to be stricter ($1.6$ as opposed to $2.0$ in the ATLAS note), although the overlap removal selection
is implemented similarly.
The mass selection used in this article is also stricter for the two photon system,
using a 2 GeV window, while the ATLAS note uses a 5 GeV window instead.
The mass selection on the two $b$-jet system is, however, stricter in the ATLAS note than in this
document, as a 60 GeV window is used here, while the ATLAS note uses a 50 GeV window.
Finally, the transverse momentum of the two $b$-jets and of the two hardest photons
have a stricter selection in this article ($150$ GeV), compared to the ATLAS one ($110$ GeV).
We also use a $\Delta \phi$ selection to reduce the impact of the backgrounds.

%%%%%%%%%%%%%%%%%%%%%%%%%%%%%%%%%%%%%%
\subsection{$hh+{\rm{jet}} \to \bar{b}b \gamma \gamma+{\rm{jet}}$}
\label{sec:analysis2}
%%%%%%%%%%%%%%%%%%%%%%%%%%%%%%%%%%%%%%
The majority of sensitivity to variations of the
Higgs trilinear coupling arises when the triangle diagram is
resonantly enhanced. Unfortunately, experimental selection cuts
often select regions of phase space far away from this regime (this is
particularly true of boosted analyses). However, this fact can be
mitigated by producing the dihiggs system at resonance in opposition
to a high $p_T$ recoiling ISR jet~\cite{us}. In this section we
therefore consider the sensitivity such an analysis would have at a
100~TeV hadron collider.

The pre-selection in this study uses a higher jet transverse momentum
and photon selection (50 GeV) and more restricted range of rapidities
($|\eta| < 2.4$) than the pre-selection cuts than the previous
section. The remaining pre-selection cuts are unchanged. However, the
event selection has been optimised for this particular topology by
demanding at least three jets and that the extra jet produced with the
$hh$ system has $p_T \geq 100$ GeV.  The two leading $b$-jets are
used as the jets from the Higgs decay. The extra jet is chosen such
that it is not one of the $b$-jets used for the Higgs reconstruction and
that it is the highest transverse momentum jet choice.
A further selection is applied on the transverse momentum
of the $\gamma\gamma$ system, which is required to be greater than
$160$ GeV.

Similar to the analysis in the previous section, the Higgs mass
requirements are applied such that $m_{bb} \in [90,150]$~GeV and
$m_{\gamma\gamma} \in [124,126]$~GeV. In this final state the impact
of the $t\bar{t}h$ background is significant, as the signal already has
extra high transverse momentum jets. To veto the impact of this
background, an upper bound is implemented on the jet multiplicity,
$N_{\textrm{jets}} \le 6$.  The signal-over-background ratio is also
slightly increased by a requirement on the $\Delta \phi$ between the
subleading $b$-jet and the $\gamma\gamma$ system, such that $\Delta
\phi(b_{2}, \gamma\gamma) > 1.6$, which can be seen to improve the
discrimination, as seen in Fig.~\ref{fig:HHj_cuts} (b).  

The final selection criterion applied is on the ratio of the
transverse momentum of the leading reconstructed Higgs and the
transverse momentum of the extra jet, which is required to be greater
than one, reducing some backgrounds as  can be seen in
Fig.~\ref{fig:HHj_cuts} (a).  The invariant masses of the
$\gamma\gamma$ and the extra jets, the 2 $b$-jets and the extra jet
and the $b\bar{b}\gamma\gamma j$ system are shown in Fig.~\ref{fig:HHj} panels (a), (b) and (c).

One important limitation is the high value of the minimum jet $p_T$,
aimed at avoiding pile up contamination and rejecting QCD
backgrounds. Signal events are rejected not only for failing the
minimum jet multiplicity requirement, but because one of the Higgs
$b$-jets may fail this selection criterion. This jet selection removes
59\% of the signal after demanding at least 2 jets only, while it also
has the side effect of rejecting an extra 50\% of the events which
fail the $b$-tagging selection when one of the $b$-jets is
rejected. The sensitivity of  the analysis could  be improved with lower transverse
momentum selection if a better photon identification performance at
low energies becomes possible in the future.

%%%%%%%%%%%%%%%%%%%%%%%%%%%%%%%%%%%%%%
\subsection*{Results}
\label{sec:results}
%%%%%%%%%%%%%%%%%%%%%%%%%%%%%%%%%%%%%%
We now combine both analyses in the $b\bar b \gamma \gamma$ channel to
formulate a constraint on the Higgs trilinear coupling in light of the
expected signal and background yields in $pp\to hh+X$ and $pp \to
hh+{\text{jet}}+X$ production. For simplicity we assume that both
measurements are statistically uncorrelated and combine them in a
binned log-likelihood hypothesis test~\cite{junk,edw}. We compute a 95\% confidence level using the CLS
method~\cite{Read:2002hq} around the SM parameter choice
$\lambda=\lambda_{\text{SM}}$ and find
{\small
\begin{equation}
  \label{eq:3ab}
  {\lambda \over \lambda_{\text{SM}}} \in \begin{cases}
    [0.672,1.406] & \hbox{no background syst.}  \\
    [0.646,1.440] & 25\%~hh, 25\%~hh+{\text{jet}} \\
    [0.642,1.448] & 25\%~hh, 50\%~hh+{\text{jet}}
  \end{cases}\,
\end{equation}
}
for an integrated luminosity of 3000/fb.  Due to the shape of
the cross section as a function of $\lambda$, there is a parameter
choice at $\lambda\simeq 4\lambda_{\text{SM}}$ with SM-like cross
sections. This region can be excluded using the high
luminosity phase of the 14 TeV LHC~\cite{Englert:2014uqa}.

In the calculation of the confidence level intervals the quoted
systematic uncertainties refer to a flat rescaling of the contributing
backgrounds. From Eq.~\eqref{eq:3ab} we can expect that a measurement
of the trilinear coupling at the 40\% level should be possible. A
$5\sigma$ discovery of the dihiggs signal will be possible with an
integrated luminosity of $700/\text{fb}$.

A number of authors have noted that a total integrated luminosity of
$3/\text{ab}$ may not be sufficient to saturate the physics potential
of a 100~TeV collider~\cite{Cohen:2014hxa,Richter:2014pga}, since the
necessary luminosity typically scales quadratically with the centre of
mass energy. We therefore also compute limits under the assumption
that $30/\text{ab}$ of data is taken. The limits shown in
Eq.~\eqref{eq:3ab} then improve to
{\small
\begin{equation}
  \label{eq:30ab}
  {\lambda \over \lambda_{\text{SM}}} \in \begin{cases}
    [0.891,1.115] & \hbox{no background syst.}  \\
    [0.882,1.126] & 25\%~hh, 25\%~hh+{\text{jet}}  \\
    [0.881,1.128] & 25\%~hh, 50\%~hh+{\text{jet}}
  \end{cases}
\end{equation}
}
in this case. We note that these limits are nearly identical to what
can be achieved with the 1~TeV luminosity upgraded ILC.

We note that that the theoretical uncertainty on the $hh$ signal
was not taken into account in the limit setting. Although
the signal theoretical uncertainty is estimated to be large
currently, mainly due to the fact that
full high order calculations are unavailable, the
100 TeV machine is relatively far in the future. It is
expected this theoretical uncertainty will be reduced in the future.

Most of the statistical pull in the $b\bar b \gamma \gamma$ channel
results from $pp \to hh+X$ production. This is expected from our
discussion in the previous section and is likely to change in other
final states such as $b\bar b \tau\tau$ \cite{us,us2}.

%%%%%%%%%%%%%%%%%%%%%%%%%%%%%%%%
\section{Discussion and Conclusions}
\label{sec:conc}
%%%%%%%%%%%%%%%%%%%%%%%%%%%%%%%%
The precision measurement of the Higgs trilinear coupling at a future
high energy hadron collider is an important motivation for the construction of
such a machine. In this paper we have performed an analysis of dihiggs
final states in the $b\bar b \gamma\gamma$ channel at a 100 TeV hadron
collider. In particular, we have explored to what extent additional hard
jet emission contributes extra statistical discriminative power. In
doing so we have implemented realistic estimates for the final state
reconstruction and arrive at the conclusion that a measurement at 40\%
level can be expected at 3/ab, which improves to the 10\% with a
factor 10 larger data set of 30/ab.

Comparing to earlier analyses performed as part of the Snowmass
process~\cite{Yao:2013ika}, we find a significantly smaller
sensitivity, which results from the more realistic treatment of
backgrounds, expected detector resolution effects, pile up and, most
importantly, fake rates.

The limiting factor of the $b\bar b \gamma \gamma$ channels is the
size of reducible backgrounds for an acceptably large signal
yield. While in this initial study we focus on the $b\bar b
\gamma\gamma$ final state, it would hence be interesting to extend
this study to other final states, and look at the use of taggers in
the $b\bar b \tau\tau$ or $b\bar b WW$ final states, which, due to
bigger signal cross sections, opens more opportunities to exploit the
high invariant mass distributions and the $hh+\text{jet}$ final state
\cite{us}.

We would like to point out that another study on the $hh\rightarrow b\bar b \gamma \gamma$
has recently appeared in \cite{contino_prep}.

\acknowledgments
CE is supported by the Institute for Particle
Physics Phenomenology Associateship programme. This research was
supported in part by the European Commission through the ‘HiggsTools’
Initial Training Network PITN-GA-2012-316704.
AJB gratefully acknowledges the support of UK Science and Technology
Facilities Council, the IPPP (Durham), and Merton College, Oxford.

%%%%%%%%%%%%%%%%%%%%%%%%%%%%%%%%%%%%%%%%%%%%%%%%%%%%%%%%%%

\end{document}